\documentclass[twocolumn,prl]{revtex4}

\usepackage{dcolumn}
\usepackage{amsmath}

\usepackage{graphicx}

\newcommand{\half}{\tfrac12}

\newcommand{\av}[1]{\left\langle#1\right\rangle}
\newcommand{\etal}{{\it{}et~al.}}
\newcommand{\defn}{\textit}

\newcommand{\mat}{\mathbf}
\renewcommand{\vec}{\mathbf}

\begin{document}

\title{Spectral community detection in sparse networks}
\author{M. E. J. Newman}
\affiliation{Department of Physics, University of Michigan, Ann Arbor, MI
  48109}
\affiliation{Santa Fe Institute, 1399 Hyde Park Road, Santa Fe, NM 87501}

\begin{abstract}
  Spectral methods based on the eigenvectors of matrices are widely used in
  the analysis of network data, particularly for community detection and
  graph partitioning.  Standard methods based on the adjacency matrix and
  related matrices, however, break down for very sparse networks, which
  includes many networks of practical interest.  As a solution to this
  problem it has been recently proposed that we focus instead on the
  spectrum of the non-backtracking matrix, an alternative matrix
  representation of a network that shows better behavior in the sparse
  limit.  Inspired by this suggestion, we here make use of a relaxation
  method to derive a spectral community detection algorithm that works well
  even in the sparse regime where other methods break down.  Interestingly,
  however, the matrix at the heart of the method, it turns out, is not
  exactly the non-backtracking matrix, but a variant of it with a somewhat
  different definition.  We study the behavior of this variant matrix for
  both artificial and real-world networks and find it to have desirable
  properties, especially in the common case of networks with broad degree
  distributions, for which it appears to have a better behaved spectrum and
  eigenvectors than the original non-backtracking matrix.
\end{abstract}

\maketitle

Since their introduction in the 1970s, spectral methods for the analysis of
large graphs and networks have become a mainstay in the study of empirical
network data~\cite{vonLuxburg07}.  One represents the structure of the
network of interest using any of several matrix forms such as the adjacency
matrix or graph Laplacian, then inspects the eigenvalues and eigenvectors
for information about network structure.  Experiments show (and it can be
proven for some model networks) that the eigenvalue spectrum typically
consists of a dense ``spectral band'' of closely spaced eigenvalues akin to
an allowed energy band in condensed matter, plus some number of outlying
eigenvalues separated from the band by a significant band gap.  The
eigenvectors corresponding to these outliers contain information about the
large-scale structure of the network and particularly about so-called
community structure, divisions of the network into groups that are tightly
knit internally but only loosely connected externally.  Spectral methods
are of particular interest because the eigenvectors can reveal large- or
global-scale structure within networks, by contrast with most other methods
of network analysis, which focus on local properties.

Spectral methods can fail, however.  The primary mode of failure is one
familiar in condensed matter physics, namely the occurrence of defects or
Griffiths singularities.  Rarely occurring but dense subgraphs within a
network can give rise to additional eigenvalues outside the main spectral
band, akin to Lifshitz tails, which can reduce or eliminate the spectral
gap.  The corresponding eigenvectors can mix with the vectors containing
community information, introducing noise or, in extreme cases, rendering
community detection impossible.  The simplest example of this phenomenon is
the occurrence of hubs in a network, vertices of unusually high degree,
which produce outlying eigenvectors strongly localized around the hubs and
lacking global structure~\cite{GKK01a,NN13}.  In networks where the average
degree of most vertices is high---much greater than one---the effect of
hubs is diluted and not usually a problem.  But in very sparse networks,
those with small average degree, the effect becomes strong and renders
conventional spectral methods virtually useless in many cases.
Unfortunately, many of the networks encountered in practical studies,
including social, technological, and biological networks, fall into this
sparse category.  The average degree of the Internet, coarse-grained at the
level of autonomous systems, for example, is about six~\cite{FFF99}.  The
average degree of metabolic networks is similar~\cite{Jeong00}.  A large
list of networks given in~\cite{Newman10} contains hardly any networks with
mean degree greater than ten.

An interesting solution to these problems has been put forward recently by
Krzakala~\etal~\cite{Krzakala13}, who propose focusing on the eigenvalues
and eigenvectors of a different matrix representation of network structure,
which they call the \defn{non-backtracking matrix}, also called the
Hashimoto edge matrix by previous authors~\cite{Hashimoto89,AFH07}.  They
show that for certain model networks this matrix displays a nonzero
spectral gap even when mean degree is very small, with outlying eigenvalues
that are well separated from the main spectral band, and hence that the
matrix avoids many of the problems that hamper other matrix representations
in the sparse limit.  Moreover, they give results, both analytic and
numerical, showing that a simple clustering of the elements of the leading
eigenvectors of the matrix is able to accurately detect community structure
in their networks.

In the present paper we propose and study a variant of the non-backtracking
matrix, which we call the \defn{flow matrix}, that shares many of the
advantages of the non-backtracking matrix, but also avoids some of its
problems.  Our study of the flow matrix is motivated by several
observations.  As we will show, the standard objective function known as
modularity, widely used in optimization schemes for community detection,
can be written straightforwardly in terms of the flow matrix, and hence
spectral methods based on the flow matrix are equivalent to approximate
optimization of the modularity, providing a connection to established
methods for the community detection problem.  Furthermore, we present
results showing that the spectrum of the flow matrix is in some respects
better behaved than that of the original non-backtracking matrix,
particularly for networks with broad degree distributions, which includes
most real-world networks.  In particular, by contrast with the
non-backtracking matrix, the flow matrix preserves a clear band edge in
such networks, and the elements of its leading eigenvectors are tightly
peaked allowing for straightforward community identification and giving
better results in some practical situations.

Consider an undirected network with $n$ vertices and $m$ edges, no
multiedges or self-loops, and only a single component.  For such a network
the original non-backtracking matrix is defined as
follows~\cite{Krzakala13,Hashimoto89}.  One first converts the network into
a directed network by replacing each undirected edge between a pair of
vertices with two directed edges pointing in opposite directions between
the same pair of vertices.  Each of the $2m$ directed edges is given a
label of the form~$i\to j$ indicating the vertex pair it connects and the
direction in which it connects them.  The non-backtracking matrix~$\mat{B}$
is a $2m\times2m$ non-symmetric matrix with one row and one column for each
directed edge and elements
\begin{equation}
B_{i\to j,k\to l} = \delta_{il}(1-\delta_{jk}),
\end{equation}
where $\delta_{ij}$ is the Kronecker delta.  In other words all elements
are zero unless edge~$i\to j$ points out of the same vertex that edge~$k\to
l$ points into, and edges~$i\to j$ and $k\to l$ are not pointing in
opposite directions between the same pair of vertices.  Note that, since
the non-backtracking matrix is not symmetric, its eigenvalues are in
general complex, unlike those of most other matrix representations for
undirected networks, but the largest eigenvalue is always real (by the
Perron--Frobenius theorem) and in some cases there may be additional
high-lying real eigenvalues as well.

The name ``non-backtracking matrix'' derives from a connection between the
matrix and the properties of non-backtracking walks.  A non-backtracking
walk~\cite{AFH07,ABLS07,FH13} is a path across the edges of a network that
is allowed to revisit a vertex visited previously but only after at least
two other vertices have been visited; immediate revisits of the form
$1\to2\to1$ are prohibited.  It is straightforward to show that powers of
the non-backtracking matrix count non-backtracking walks and that traces of
powers count closed non-backtracking walks.  The spectrum of any matrix is
given entirely by such traces of powers, via a derivative of the Stieltjes
transform~\cite{Tao12}, and hence in this case by counts of closed
non-backtracking walks.

Note that any subgraph of the network that takes the form of a tree,
attached to the rest of the network at only a single point, contains no
non-backtracking walks, since all such walks contain at least one loop and
a tree contains none.  Hence the presence (or absence) of such trees in the
network has no effect on the spectrum and one can remove them.  In the
developments that follow we will assume that all such dangling trees have
been removed, which will make our calculations simpler.  A network with its
dangling trees removed is called a \defn{2-core}.  We will revisit the
question of dangling trees in networks toward the end of this paper.

The primary result of Krzakala~\etal~\cite{Krzakala13} is that for certain
classes of model networks the non-backtracking matrix has a sharp edge to
its spectral band, with no Lifshitz tails, and a nonzero spectral gap,
regardless of the average degree.  It thus appears the matrix is immune to
the problems that plague the Laplacian and other graph representations in
the low-degree limit.  Krzakala~\etal\ give arguments indicating that the
complex eigenvalues of the non-backtracking matrix for random graphs with
Poisson degree distribution should lie with a circle of radius
$\sqrt{\av{d}}$ in the complex plane, where $\av{d}$ is the mean degree of
the graph, and in practice this result seems to work well on simulated
networks.  They also give a generalization to networks that have
non-Poisson degrees, as most real-world networks do.  For such networks the
mean degree~$\av{d}$ is replaced with the mean expansion rate of the
network, which is $\av{d^2}/\av{d}-1$ for uncorrelated networks.  This
generalization, however, may be less useful in practical situations.  In
particular, in the common case of networks with power-law degree
distributions the expansion rate diverges and with it the bound on the
eigenvalues, and even for broad degree distributions with finite moments
the bound may be very high.

In this paper we study a variant of the non-backtracking matrix which
appears to be better behaved when networks have broad degree distributions,
and which also has close connections to previously studied methods of
community detection that are known to perform reliably under realistic
circumstances.  The matrix we study, which we call the flow matrix, is
defined on the same $2m$ directed edges as the non-backtracking matrix, and
has elements
\begin{equation}
F_{i\to j,k\to l} = {\delta_{il}(1-\delta_{jk})\over d_i-1},
\end{equation}
where $d_i$ is the degree of vertex~$i$.  The flow matrix can be thought of
as a conservative-flow version of the non-backtracking matrix that
describes the motion of a conserved quantity around the network, but
subject to a non-backtracking constraint.  Its powers count
non-backtracking walks weighted inversely by the product $\prod_i (d_i-1)$
for the vertices they traverse.

The flow matrix has a close connection to the standard objective function
known as modularity, as we can show by the following argument.  Given a
network and a division of that network into communities or groups, the
modularity~$Q$ is defined to be the fraction of edges that fall within
communities minus the expected value of the same fraction if edges were
placed at random~\cite{NG04}.  In mathematical terms,
\begin{equation}
Q = {1\over2m} \sum_{ij} \biggl[ A_{ij} - {d_i d_j\over2m} \biggr]
    \delta_{g_ig_j},
\label{eq:modularity1}
\end{equation}
where $g_i$ is the label of the group to which vertex~$i$ belongs and
$A_{ij}$ is an element of the adjacency matrix~$\mat{A}$, having value 1 if
there is an edge between vertices $i$ and~$j$ and zero otherwise.  The
modularity quantifies how good our division of the network is and
modularity-based community detection methods find good divisions by
maximizing it over divisions, i.e.,~over the group assignments~$g_i$.

Consider the simplest example of this maximization problem, where the
network is divided into just two groups (of any size) and define a set
of~$n$ index variables~$s_i$, one for each vertex, such that $s_i=+1$ if
vertex~$i$ belongs to group~1 and $-1$ if it belongs to group~2.  Then
consider the following scalar quadratic form involving the flow matrix:
\begin{equation}
R = \vec{u}^T(\mat{F}-\vec{1}\vec{1}^T)\vec{v},
\end{equation}
where $\vec{u}$ and $\vec{v}$ are two $2m$-element vectors that we choose
and $\vec{1}$ is the uniform unit vector $\vec{1} =
(1,1,1,\ldots)/\sqrt{2m}$.  If we make the particular choice $u_{i\to j} =
v_{i\to j} = s_j$, meaning that the elements of both vectors are equal to
the group index of the vertex to which the corresponding edge points, then
\begin{align}
\vec{u}^T\mat{F}\vec{v} &= \sum_{\substack{\textrm{edges\ } i\to j\\
  \textrm{edges\ } k\to l}} {\delta_{il}(1-\delta_{jk})\over d_i-1}\, s_j s_l
  \nonumber\\
  &= \sum_i {s_i\over d_i-1} \sum_{jk} A_{ij} A_{ki}
     (1-\delta_{jk}) s_j \nonumber\\
  &= \sum_i {s_i\over d_i-1} \sum_{j} (d_i - 1) A_{ij} s_j
   = \vec{s}^T\mat{A}\vec{s},
\label{eq:deriv1}
\end{align}
where $\vec{s}$ is the $n$-element vector with elements~$s_i$.  Also
\begin{align}
\vec{u}^T\vec{1}\vec{1}^T\vec{v} &=
   {1\over2m} \sum_{\substack{\textrm{edges\ } i\to j\\
   \textrm{edges\ } k\to l}} s_j s_l
   = {1\over2m} \sum_{ijkl} A_{ij} A_{kl} s_j s_l \nonumber\\
  &= {1\over2m} \sum_{jl} d_j d_l s_j s_l
   = \vec{s}^T {\vec{d}\vec{d}^T\over2m} \vec{s},
\label{eq:deriv2}
\end{align}
where $\vec{d}$ is the $n$-element vector with elements equal to the
degrees~$d_i$.  Noticing that $\half(s_is_j+1)$ is 1 if $i$ and~$j$ are in
the same group and zero otherwise, we can now combine
Eqs.~\eqref{eq:modularity1}, \eqref{eq:deriv1}, and~\eqref{eq:deriv2} to
get an expression for the modularity:
\begin{align}
Q &= {1\over4m} \sum_{ij} \biggl[ A_{ij} - {d_i d_j\over2m} \biggr]
                (s_i s_j + 1)
   = {1\over4m} \vec{s}^T \biggl[ \mat{A} - {\vec{d}\vec{d}^T\over2m}
     \biggr] \vec{s} \nonumber\\
  &= {1\over4m} \vec{u}^T \bigl( \mat{F} - \vec{1}\vec{1}^T \bigr) \vec{v}
   = R,
\label{eq:modularity2}
\end{align}
where in the second equality we have made use of the fact that $\sum_{ij}
A_{ij} = \sum_i d_i = 2m$.

In other words, $R$~is simply equal to the modularity.

The goal of modularity-based community detection is to find the group
indices~$s_i$ that maximize the modularity.  This optimization is known, in
general, to be a hard computational problem~\cite{Brandes07}, but good
approximate solutions can be found by a variety of heuristics.  Here we use
a standard relaxation technique, in which we relax the condition that the
elements of $\vec{u}$ and~$\vec{v}$ are equal to~$s_i=\pm1$, allowing them
to take any real values.  We must be careful, however---if the magnitudes
of $\vec{u}$ and~$\vec{v}$ can become arbitrarily large then there is no
limit on the value of the modularity.  To prevent this happening we impose
an additional constraint.  We note that when $u_{i\to j} = v_{i\to j} =
s_j$ we have $\vec{u}^T\vec{v}=2m$, a constant, and we will impose the same
constraint on our relaxed optimization.  Clearly the original unrelaxed
values of the vector elements satisfy this constraint, so the relaxed
optimization includes these values, but it includes many other values as
well.

The maximization of Eq.~\eqref{eq:modularity2} in this relaxed solution
space is straightforward.  Introducing a Lagrange multiplier~$\lambda$ to
enforce the constraint on~$\vec{v}$ and differentiating with respect to the
elements of~$\vec{u}$, we find that the maximum modularity occurs when
\begin{equation}
\bigl( \mat{F} - \vec{1}\vec{1}^T \bigr) \vec{v} = \lambda \vec{v}.
\end{equation}
In other words~$\vec{v}$ is an eigenvector of the matrix~$\mat{F} -
\vec{1}\vec{1}^T$ with eigenvalue~$\lambda$.  Substituting this solution
back into Eq.~\eqref{eq:modularity2} then gives
\begin{equation}
Q = {1\over4m} \vec{u}^T \bigl( \mat{F} - \vec{1}\vec{1}^T \bigr) \vec{v}
  = {\lambda\over4m} \vec{u}^T\vec{v} = {\lambda\over2}.
\label{eq:modularity3}
\end{equation}
Thus the maximum modularity is given by setting $\vec{v}$ equal to the
eigenvector for the largest (most positive) eigenvalue of the matrix.

This constitutes an exact solution of the relaxed maximization problem,
which we now use as a guide to the solution of the original unrelaxed
problem of finding the quantities~$s_i=\pm1$ on the vertices.  Recalling
that in the unrelaxed formulation all vector elements~$v_{i\to j}$ are
equal to~$s_j$, we see that our relaxed solution for~$\vec{v}$ in fact
gives us $d_j$ different values for~$s_j$, one for every edge incident
on~$j$.  We estimate the true~$s_j$ by taking the average of these values
and rounding to the nearest~$\pm1$, which in practice means $s_j=+1$ if the
sum $\sum_i A_{ij} v_{i\to j}$ is positive and $s_j=-1$ if the sum is
negative.

Thus our spectral algorithm is a simple one: we calculate the leading
eigenvector of the matrix~$\mat{F} - \vec{1}\vec{1}^T$, sum the elements
pointing to each vertex, and divide the vertices into two groups according
to the signs of these sums.  This is essentially the algorithm used by
Krzakala~\etal\ in their calculations except that we use the normalized
version of the non-backtracking matrix---the flow matrix---rather than the
original non-backtracking matrix.  As with all relaxation methods, this one
gives only an approximation to the true optimum of our objective function,
but as we will see it performs well in practice.

We observe the following about the eigenvectors of the matrix~$\mat{F}$.
The $i\to j$ row of the matrix contains~$d_i-1$ nonzero elements with
value~$1/(d_i-1)$ each, meaning that the uniform vector~$\vec{1}$ is a
(properly normalized) eigenvector of the matrix with eigenvalue~1.  Thus it
is also an eigenvector of $\mat{F} - \vec{1}\vec{1}^T$, but with eigenvalue
zero.  All other eigenvalues and eigenvectors of $\mat{F} -
\vec{1}\vec{1}^T$ are the same as those for~$\mat{F}$.  By the
Perron--Frobenius theorem the vector~$\vec{1}$ is the leading eigenvector
of~$\mat{F}$ (since it has all elements positive and the network is
connected), while for the matrix~$\mat{F} - \vec{1}\vec{1}^T$, which has
this eigenvector removed, the most positive eigenvalue must equal the
second eigenvalue of~$\mat{F}$ (so long as $\mat{F} - \vec{1}\vec{1}^T$ has
any positive eigenvalues).  Thus we can, as we wish, perform community
detection using either the leading eigenvector of $\mat{F} -
\vec{1}\vec{1}^T$ or the second eigenvector of~$\mat{F}$.  The results will
be identical.  In this paper we do the latter, since it is slightly simpler
in practice.

\begin{figure}
\begin{center}
\includegraphics[width=\columnwidth]{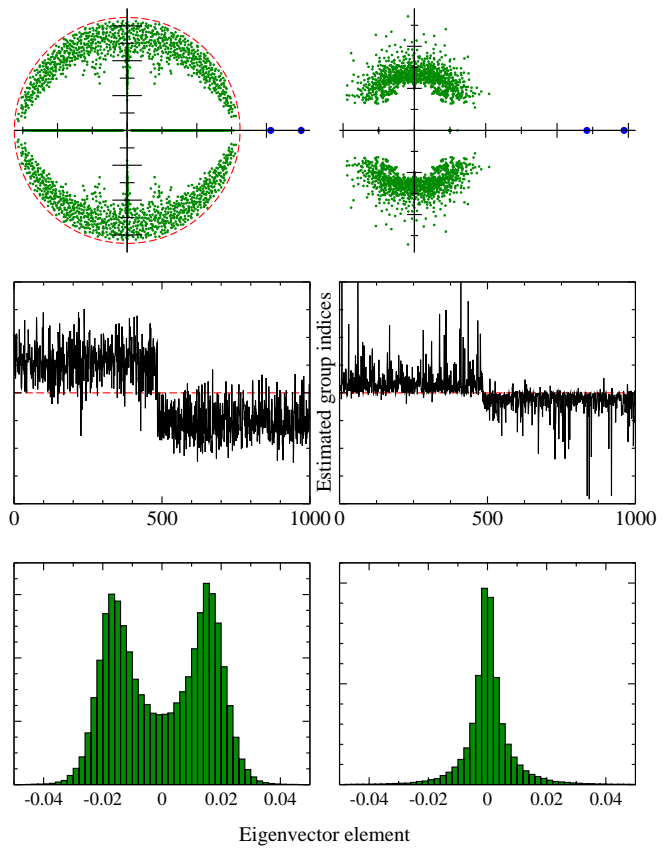}
\end{center}
\caption{Complex spectra, estimated group indices~$s_i$, and distribution
  of eigenvector elements for a 1000-node network generated using the
  degree-corrected stochastic block model with a power-law degree
  distribution with exponent~$-2.5$.  The left three panels are for the
  flow matrix~$\mat{F}$ described in this paper; the right three panels are
  for the non-backtracking matrix of~\cite{Krzakala13}.  Each dot in the
  top two panels represents one eigenvalue, plotted in the complex plane.
  The vertical scales on the center two panels are the same.  The
  histograms in the two bottom panels are each averaged over 100 model
  networks with the same parameters.}
\label{fig:dcsbm}
\end{figure}

We have performed a number of numerical calculations of the spectrum of the
flow matrix, along with tests of the community detection algorithm derived
above.  Figure~\ref{fig:dcsbm} shows results for (the largest 2-core of) a
computer-generated network created using the degree-corrected block model
of~\cite{KN11a} with two communities.  This model allows one to generate
networks that simultaneously contain planted community structure and
nontrivial degree distribution.  In our tests we generated networks with a
power-law degree distribution with exponent~$-2.5$ to test the behavior of
our methods in the case of realistically high degree variance.  The top
left panel of Fig.~\ref{fig:dcsbm} shows the spectrum of the flow matrix in
the complex plane for a network of $n=1000$ vertices.  As the figure shows,
the eigenvalues fall neatly within a circular region with a clear band edge
and there are two outlying eigenvalues.  An argument analogous to that
of~\cite{Krzakala13} indicates that the eigenvalues lie within a circle of
radius $\sqrt{\av{d/(d-1)}/\av{d}}$, which is never greater than~1.  Such a
circle is plotted in Fig.~\ref{fig:dcsbm} and appears to agree well with
the numerical results.

Of the two outlying eigenvalues, the higher one has value~1 and
eigenvector~$\vec{1}$ as we have said; the lower one contains the community
structure.  The center left panel shows our relaxed estimates of the group
membership variables~$s_i$ from the second eigenvector, and the two planted
groups in the network (of equal sizes in this case) are clearly visible.
Dividing the vertices according to the signs classifies 93\% of them into
the correct groups.  Apart from modest statistical fluctuations, these
results appear robust over repetitions of the experiment with the same
parameters.  The bottom left panel of the figure shows the distribution of
the eigenvector elements for the flow matrix, with two clear peaks that
correspond closely to the planted communities in the network.

For comparison we show in the right three panels of the figure the
corresponding quantities for the original non-backtracking matrix.  The
non-backtracking matrix also does a good job of classifying vertices into
groups, with only a modestly lower fraction 90\% of vertices classified
correctly.  The spectrum of the matrix is less well behaved, however.  As
the top right panel shows, the spectrum is more diffuse than that of the
flow matrix, having no clear circular edge.  And while there are still two
outlying eigenvalues, the estimates of~$s_i$ calculated from the second
eigenvector, shown in the center right panel, are noisier, with a strongly
non-Gaussian distribution and occasional large fluctuations.  Similar
fluctuations are also seen in, for example, spectra of the ordinary graph
Laplacian matrix for networks with broad degree distributions, and are
known to give rise to poor performance for algorithms based on those
spectra.  The bottom right panel shows the distribution of eigenvector
elements for the non-backtracking matrix, and the bimodal distribution seen
in the flow matrix is gone, replaced by just a single peak with no clear
separation between the communities.

\begin{figure}
\begin{center}
\includegraphics[width=\columnwidth]{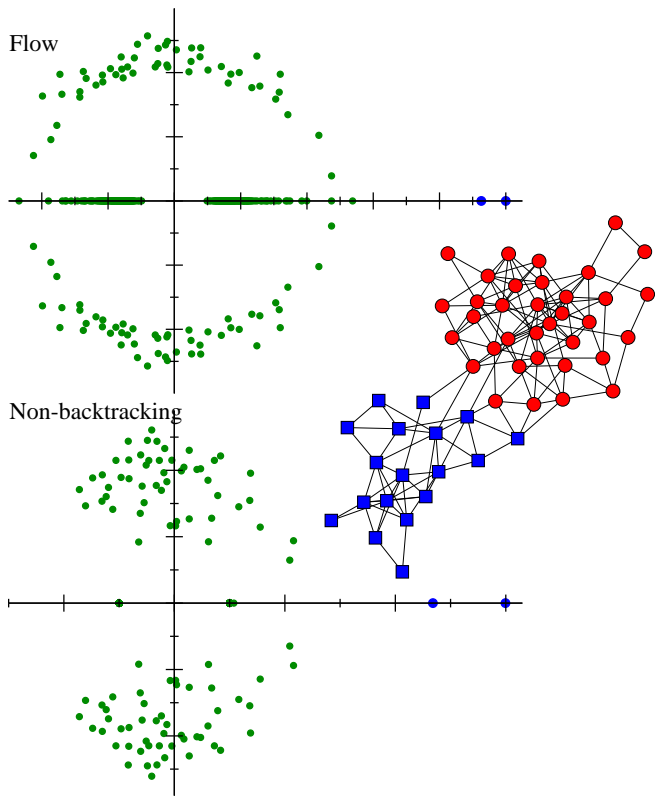}
\end{center}
\caption{Complex spectra and community structure for the social network of
  bottlenose dolphins of Lusseau~\etal~\cite{Lusseau03a}.  Top: spectrum of
  the flow matrix.  Bottom: spectrum of the non-backtracking matrix.
  Right: division into two groups found using the second eigenvector of the
  flow matrix.  The division found corresponds closely to that observed in
  the original study.}
\label{fig:dolphins}
\end{figure}

A similar pattern is seen in applications to real-world networks.
Figure~\ref{fig:dolphins}, for example, shows an analysis of an animal
social network, a network of bottlenose dolphins studied previously by
Lusseau~\etal~\cite{Lusseau03a}.  This network is believed to divide into
two clear communities and the spectrum of the flow matrix confirms this,
with a compact circular spectral band and two outlying eigenvalues (top
graph in figure).  The split derived from the second eigenvector (center
right in the figure) corresponds closely to that identified in the original
study of the dolphin community.  The non-backtracking matrix also does a
good job of revealing the community structure, but its spectrum is once
again more diffuse (bottom).

Before closing, let us return to an issue raised at the start of this
paper, that of dangling trees attached to a network.  The original
non-backtracking matrix takes no account of such trees---when they are
removed, leaving the 2-core of the network, both the eigenvalues and
eigenvectors of the matrix are unchanged (within the 2-core) from those of
the full network.  This is perhaps the most serious drawback of the method
as proposed---widely used and trusted methods of community detection, such
as modularity maximization or inference methods based on block models, can
give very different answers for networks with and without dangling trees,
and hence can disagree strongly with methods based on the non-backtracking
matrix.  The variant matrix discussed here to some extent remedies this
problem: its spectrum does change when dangling trees are added to or
removed from the network, and one can write a version of
Eq.~\eqref{eq:modularity2} that is correct for all networks, not just those
consisting of a 2-core, so that relaxations give an approximation to the
maximum modularity partitioning of any network.  The problem is not
completely solved, however, as one can see by considering the extreme case
of a network composed of a single tree with no 2-core at all.  Because the
spectra of both the matrices~$\mat{B}$ and~$\mat{F}$ are determined by
counts of non-backtracking walks, and because there are no such walks on a
tree, all eigenvalues of both matrices are zero and the eigenvectors fail
to give any partition for such a network, even though other methods,
including exact modularity maximization, give sensible partitions when
applied to trees.  It is an open question whether and how this problem can
be remedied.

The author thanks Tammy Kolda, Cris Moore, Elchanan Mossel, Joe Neeman, and
Lenka Zdeborov\'a for useful conversations.  This work was funded in part
by the National Science Foundation under grant DMS--1107796 and by the Air
Force Office of Scientific Research (AFOSR) and the Defense Advanced
Research Projects Agency (DARPA) under grant FA9550--12--1--0432.

\end{document}